\documentclass[twocolumn,showpacs,showkeys]{revtex4}
\usepackage{amsfonts}
\usepackage[T2A]{fontenc}
\usepackage[cp1251]{inputenc}
\usepackage{amsmath}
\usepackage{amssymb}
\usepackage[english]{babel}
\usepackage{graphicx}

\begin{document}

\title{On the radiative lifetime of free-moving two-dimensional excitons}

\author{A. V. Paraskevov}

\affiliation{National Research Centre "Kurchatov Institute", Kurchatov Sq. 1, Moscow 123182, Russia}

\begin{abstract}

A simple microscopic mechanism explaining the linear dependence of
the radiative lifetime of free-moving two-dimensional excitons on
their effective temperature is suggested. It is shown that there
exists a characteristic effective temperature (of about few Kelvin)
defined by the exciton-acoustic phonon interaction at which the
radiative lifetime is minimal. Below this temperature the lifetime
starts to increase with decreasing temperature. The correspondence
with previous theoretical and experimental results is discussed.

\end{abstract}



\maketitle

\section{Introduction}The mechanism of optical recombination of free-moving two-dimensional
(2D) excitons in semiconductor quantum wells (QWs) is an intriguing
open question in the field of exciton physics, especially, in the
light of recent experiments on spatially-resolved exciton
luminescence indicating non-trivial diffusion of excitons in the QW
plane \cite{But}. In fact, optical decay of a free-moving 2D exciton
leads to the formation of either a three-dimensional (3D) photon or
a quasi-2D exciton polariton, if the exciton system is placed in a
resonant planar microcavity \cite{Agr66, Hanam, Andr, Citr, Litt,
Ivanov}. The decay can depend on interactions of excitons with
acoustic phonons and defects. Indeed, the presence of a "third body"
(e.g., 3D acoustic phonons) could open a fast channel of radiative
decay similar to three-body recombination in plasmas and
cold gases. In what follows, we study the influence of
exciton-acoustic phonon interaction on optical recombination of
free-moving excitons in a narrow QW, without any resonant
cavity.

The first phenomenological theory of optical decay of free-moving QW
excitons was suggested in Ref.\cite{Feld}. It was assumed that the
excitons had a Maxwellian distribution over momenta with some effective
temperature $T$. The influence of the exciton-phonon interaction on
exciton radiative lifetime $\tau_{R}$ was taken into account by
sharing the oscillator strength of an exciton with zero momentum
\textit{equally} among all momentum states within a phenomenological
spectral width. It was found that $\tau_{R}\propto T$. This result
was justified by the reasonable agreement with experimental data for
the case of excitons nearly thermalized to the lattice temperature $T_{0}$, i.e., at $T\approx T_{0}$.
From the physical point of view this means that
the increase of $\tau_{R}$ with $T_{0}$ is due to the fact that the
scattering of excitons on phonons can lead to a decrease of the
fraction of excitons (in momentum space) that are able to decay
radiatively.

In Refs.\cite{Hanam, Andr, Litt} the optical recombination rate
$\Gamma(p)$ of an exciton with in-plane momentum $p$ was derived
assuming that only the exciton with momentum $p<k=\omega_{0}/c$ can decay
radiatively, where $k$ and $\omega_{0}$ are wave vector and frequency of a 3D
photon, respectively, and $c$ is the speed of light in the medium. The averaging of $\Gamma(p)$ over the thermal
distribution of excitons gave the previous result
$\tau_{R}\propto T$ \cite{Andr, Litt} (see Appendix). However, if one considers the
case when a 2D exciton optically decays with the simultaneous
emission of a 3D acoustic phonon, then the condition $p<k$ for the
radiative decay is apparently invalid even at $T_{0}=0$.

In this paper we study the influence of exciton-acoustic phonon
interaction on optical decay of free-moving QW excitons for the case
when the exciton kinetic energy is small enough to neglect optical
phonon emission. In particular, starting from the premise that
free-moving 2D exciton decays only with the formation of a 3D photon
and a 3D acoustic phonon (or, equivalently, that the exciton decays just
after scattering from the phonon with some definite momentum), we
have derived the dependence $\tau_{R}\propto T$ at high $T$, but in
completely different way compared to Ref.\cite{Andr}. In
addition, our model gives an intuitively-expected upturn of the
dependence at very low $T$. Note that the above premise is based on two general reasons: (i) an exciton is a metastable system so any
additional interaction would likely facilitate its decay; (ii) the
exciton-acoustic phonon interaction is something unavoidable even
in the case of zero lattice temperature, where excitons can only emit
phonons.

\section{Model}Let us consider a quasi-equilibrium system of
free-moving 2D excitons with effective temperature $T$, where
excitons can interact with 3D longitudinal acoustic (LA) phonons. Then exciton distribution function $f_{p}$ over quasi-momentum
$\mathbf{p}$ is Maxwellian, $f_{p}=\exp\left(
-(E_{p}-\mu)/T\right)  $, where dispersion law $E_{p}=p^{2}/2m$, $m$
is the exciton effective mass, and $\mu<0$ corresponds to the
chemical potential of the exciton system, $\left\vert \mu\right\vert
\gg T$. The exciton density is
$n=\frac{1}{S}%
{\displaystyle\sum\limits_{\mathbf{p}}}
f_{p}=mT\exp\left(  \mu/T\right)  /\left(  2\pi\hbar^{2}\right)  $,
where $S$ is the X-Y plane area. For clarity we set lattice
temperature $T_{0}=0$, so the excitons can only emit phonons. In
what follows we use the notation: $\mathbf{p}$ is 2D exciton
momentum in X-Y plane, $\mathbf{k}$ is 3D photon wave vector
($\left\vert \mathbf{k}\right\vert =\omega_{0}/c$) and
$\mathbf{q}=\left(  \mathbf{q}_{\parallel},q_{z}\right)  $ is 3D
acoustic phonon wave vector. By default we omit $\hbar$ in most
formulae for phonon and photon momenta.

\begin{figure}[!t]
\includegraphics[width=8.7cm]{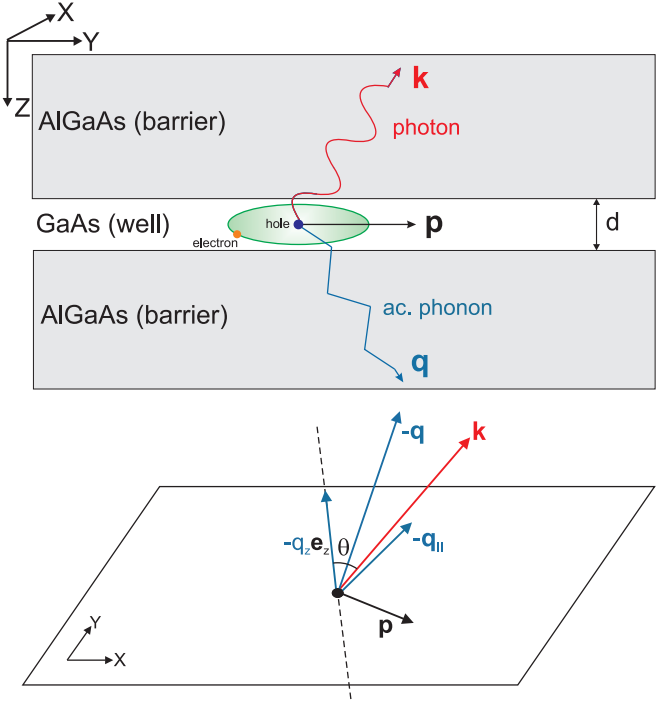}\caption{Top: Free-moving exciton
(bound electron-hole pair, shown by ellipse) in a GaAs/AlGaAs
quantum well. "2D exciton" means that the well width $d\ll a_{B}$,
where $a_{B}$ is exciton Bohr radius in GaAs. Bottom: Schematic of
the momentum conservation law $\mathbf{p}=\mathbf{k+q}$:
$\mathbf{p}$ is 2D exciton momentum in X-Y plane, $\mathbf{k}$ is 3D
photon wave vector and $\mathbf{q}=\left(
\mathbf{q}_{\parallel},q_{z}\right)$ is 3D acoustic phonon wave vector.}%
\label{Fig1}%
\end{figure}

Let us suppose that a free-moving 2D exciton can emit a 3D photon only
if $\mathbf{p}=\mathbf{k+q}$, or (Fig. \ref{Fig1})
\begin{equation}
\left(  \mathbf{p}-\mathbf{q}_{\parallel}\right)
^{2}=\mathbf{k}^{2}\sin ^{2}\theta,\text{ }q_{z}=-k_{z},\text{
}\cos\theta\equiv k_{z}/\left\vert \mathbf{k}\right\vert
.\label{condop}%
\end{equation}
If $\mathbf{q}=\mathbf{0}$ the momentum conservation law
(\ref{condop}) would allow only the formation of a 2D
exciton-polariton in the X-Y plane so the actual optical decay would be
prohibited.

Further, let us assume that an exciton recombines \textit{instantly}
after scattering from an appropriate phonon (\ref{condop}). In
particular, an exciton with initial 2D momentum $\mathbf{p}$ emits a
3D phonon with wave vector $\mathbf{q}$ so that after emission the
final 2D momentum of the exciton is
$\mathbf{p}-\mathbf{q}_{\Vert}$. The transition rate for such a process is given by
\begin{equation}
w(\mathbf{p},\mathbf{q})=\frac{2\pi}{\hbar}\left\vert M_{q}\right\vert
^{2}f_{p}\left(  1+f_{p-q_{\Vert}}\right)  \delta\left(  E_{p}-E_{p-q_{\Vert}%
}-\hbar sq\right)  \label{W}%
\end{equation}
with $q=\sqrt{q_{\Vert}^{2}+q_{z}^{2}}$. (Here it is implied that
the excitons are bosons and the lattice temperature $T_{0}=0$.) Note
that for pure 2D exciton - 3D phonon scattering, when the
exciton does not decay, due to the quantum confinement the value of
$q_{z}$ is undefined and one should average over it in (\ref{W}).

The matrix element $M_{q}$ for the exciton-LA phonon interaction in
QWs due to the deformation potential was derived in
Ref.\cite{Takag}. For simplicity, we consider the limiting case of
zero QW width, i.e., a 2D layer. If $q_{\Vert}a_{B}\ll1$, where
$a_{B}$ is the 2D-exciton Bohr radius, $\left\vert M_{q}\right\vert
^{2}$ can be reduced to
\begin{equation}
\left\vert M_{q}\right\vert ^{2}\approx M^{2}a_{B}\left(  q_{\Vert}^{2}%
+q_{z}^{2}\right)  ^{1/2},\label{Mq}%
\end{equation}
where $M^{2}=\frac{\hbar\left(  D_{c}-D_{v}\right)  ^{2}}{4\pi\rho
Vsa_{B}}$, $\rho$ is the crystal density, $V$ is its volume,
$D_{c(v)}$ is the deformation potential for an electron in the conductance
(valence) band, and $s$ is the sound velocity. For definiteness, in all numerical estimates we consider
heavy-hole excitons in GaAs. For this material one has
$a_{B}\approx10^{-6}$ cm, $m\approx0.06m_{e}$ ($m_{e}$ is free
electron mass), $s\approx5\cdot10^{5}$ cm/s, $\rho=5.3$ g/cm$^{3}$
and $\left\vert D_{c}-D_{v}\right\vert \approx10$ eV \cite{Hanam}.

Since an exciton has a finite size and internal structure, there are
two conditions required to treat the excitons as point-like
structureless particles. (I) The de Broglie wavelength of an exciton
should be much larger than the exciton Bohr radius,
$\lambda_{dB}=\hbar/\sqrt{2mT}>>a_{B}$. This sets the upper limit
for the effective exciton temperature, $T<<E_{b}$, where
$E_{b}=2\hbar^{2}/\left(  ma_{B}^{2}\right)\approx3\cdot10^{2}$ K is
the binding energy of a 2D exciton. Note that
$\lambda_{dB}^{-1}\approx10^{6}$ cm$^{-1}$ at $T=10^{2}$ K and
$\lambda_{dB}^{-1}\approx3\cdot10^{5}$ cm$^{-1}$ at $T=10$ K. In
addition, since we consider non-degenerate excitons,
$T>>T_{c}=2\pi\hbar^{2}n/m$. Hence the theory could be applied at $1$ K
$\lesssim T\lesssim50$ K with exciton density $n\lesssim10^{10}$
cm$^{-2}$. (II) To exclude the influence of the exciton internal
structure on the exciton-acoustic phonon interaction one needs to assume
that $\hbar^{2}/\left(  ma_{B}%
^{2}\right)  \gg\hbar sq$ or $qa_{B}\ll\beta^{-1}$, where $\beta
=a_{B}(ms/\hbar)\approx2\cdot10^{-2}$. We generalize this condition
to $qa_{B}\lesssim1$ to use both the matrix element $M_{q}$ in
simplified form (\ref{Mq}) and the fact that the photon wave vector
$k\sim10^{5}$ cm$^{-1}$ \cite{But} with $ka_{B}\sim0.1$.

\section{Radiative lifetime and luminescence intensity}From
condition (\ref{condop}) it follows that for given $\mathbf{p}$ and
$\mathbf{k}$ there exists only one phonon with wave vector
$\mathbf{q}$ defined by (\ref{condop}), after the emission of which
the exciton can decay radiatively. Then the average radiative
lifetime $\tau_{R}$ of an exciton in the system can be determined
through%
\begin{equation}
\frac{n}{\tau_{R}}=\frac{1}{S}%
{\displaystyle\int\limits_{0}^{\pi/2}}
\frac{d\theta}{\left(  \pi/2\right)  }%
{\displaystyle\sum\limits_{\mathbf{p},\mathbf{\mathbf{q}_{\parallel}}}}
w(\mathbf{p},(\mathbf{q}_{\parallel},\left\vert \mathbf{k}\right\vert
\cos\theta))\delta_{(\mathbf{p}-\mathbf{q}_{\parallel})^{2},\mathbf{k}^{2}%
\sin^{2}\theta}.\label{taur}%
\end{equation}
Here $\delta_{p,p^{\prime}}=1$ if $p=p^{\prime}$ and
$\delta_{p,p^{\prime}}=0$ otherwise. On the other hand, one can
determine the optical recombination rate $\Gamma(p)$ of the exciton
with momentum $\mathbf{p}$ from the relation
\begin{equation}
n/\tau_{R}=\frac{1}{S}%
{\displaystyle\sum\limits_{\mathbf{p}}}
\Gamma(p)f_{p}=I,\label{Int}%
\end{equation}
where $I$ is the average intensity of the exciton luminescence.

Substituting (\ref{W}) and (\ref{Mq}) in Eqs.(\ref{taur}-\ref{Int}) one obtains%
\begin{equation}
\Gamma(p)=\frac{4a_{B}M^{2}}{\hbar}%
{\displaystyle\int\limits_{0}^{\pi/2}}
d\theta%
{\displaystyle\sum\limits_{\mathbf{\mathbf{q}_{\parallel}}}}
q_{\ast}\delta\left(  \frac{p^{2}}{2m}-\frac{\hbar^{2}k^{2}}{2m}\sin^{2}%
\theta-\hbar sq_{\ast}\right)  ,\label{Gam}%
\end{equation}
with $q_{\ast}=\sqrt{q_{\Vert}^{2}+k^{2}\cos^{2}\theta}$.

\begin{figure}[!t]
\includegraphics[width=8.7cm]{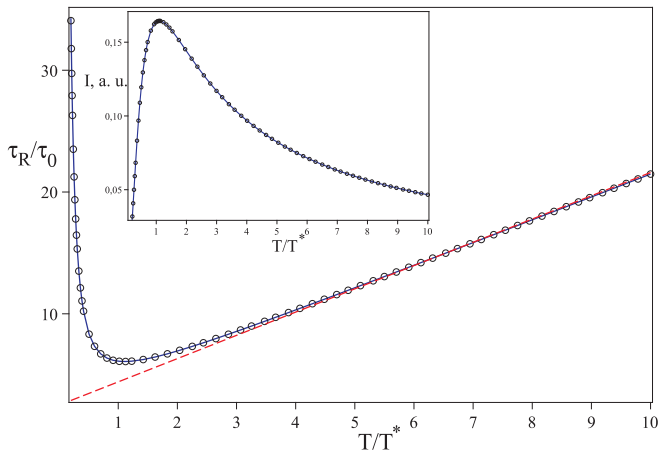}\caption{Radiative exciton lifetime $\tau_{R}$
as a function (\ref{lifetime}) of the effective temperature $T$ in
the exciton system (lattice temperature $T_{0}=0$). One can see the
minimum at $T$ = $T^{\ast}$. Inset: luminescence intensity
(\ref{intensity}) vs $T$. Solid lines are analytical curves given by
Eqs. (\ref{intensity}), (\ref{lifetime}) with $J\left(x\right)$ in
form (\ref{Jint-ap}), circles are numerical evaluation of Eqs.
(\ref{intensity}), (\ref{lifetime}) with $J\left(x\right)$ given by
(\ref{Jint}), and the dash line is
$1.91(T/T^{\ast})+2.5$.}%
\label{Fig2}%
\end{figure}

Evaluating the sums in (\ref{Int}) with the use of the condition $q_{\Vert}%
a_{B}\lesssim1$, one gets (inset in Fig. \ref{Fig2})%
\begin{equation}
I=\frac{1}{S}%
{\displaystyle\sum\limits_{p}}
\Gamma(p)f_{p}\approx\frac{n}{\tau_{0}}x^{2}J\left(  x\right)
,\label{intensity}%
\end{equation}
Here $x=T/T^{\ast}$, $T^{\ast}=\hbar s/a_{B}\approx3.6$ K, $\tau_{0}^{-1}%
=2M^{2}S/\left(  \hbar^{2}sa_{B}\right)$, where $\tau_{0}$ is
the density-and-temperature independent time constant, and
\begin{equation}
J\left(  x\right)  =%
{\displaystyle\int\limits_{0}^{\pi/2}}
d\theta\exp\left(-\frac{\alpha^{2}\sin^{2}\theta}{2\beta x}\right)
{\displaystyle\int\limits_{u/x}^{\frac{1}{x}\sqrt{1+u^{2}}}}
y^{2}dy\exp\left(  -y\right)  \label{Jint}%
\end{equation}
with $\alpha=ka_{B}\sim0.1$ and $u=\alpha\cos\theta$. In the pure 2D
case (i.e., for a QW with zero width) one has to consider $\tau_{0}$
as a phenomenological parameter.

According to (\ref{Int}), the temperature dependence of the radiative
exciton lifetime is given by (Fig. \ref{Fig2})
\begin{equation}
\tau_{R}=\tau_{0}/\left(  x^{2}J\left(  x\right)  \right)  .\label{lifetime}%
\end{equation}
It is seen that at large temperature $T\gtrsim T^{\ast}$ the classical
dependence $\tau_{R}\propto T$ is reproduced. Decreasing temperature
below $T^{\ast}$, one finds the intuitively expected increase of
$\tau_{R}$. Note that for GaAs at $T\lesssim1$ K the contribution of
piezoelectric coupling in the exciton-acoustic phonon interaction
becomes significant \cite{add} so that one can expect smoother
dependence $\tau_{R}(T)$ in this region.

Calculating the integral over $y$ in (\ref{Jint}) and expanding
$\exp\left( -\frac{\alpha}{x}\cos\theta\right)$ in the resulting
expression in powers of $\alpha$ up to and including the cubic terms, then
integrating over $\theta$ one gets an approximate evaluation of $J\left(  x\right)$ in the form%
\begin{equation}
J\left(  x\right) \approx F(x)G_{0}\left(  \xi\right)  +\frac{\alpha^{2}%
}{2x^{3}}\exp(-1/x)G_{2}\left(  \xi\right)  -\frac{\alpha^{3}}{3x^{3}}%
G_{3}\left(  \xi\right)  ,\label{Jint-ap}%
\end{equation}
which coincides almost exactly with the numerical solution (Fig.
\ref{Fig2}). Here $\xi
=\alpha^{2}/\left(  2\beta x\right) $ and%
\begin{align*}
F(x)  & =2-\left(2+\frac{2}{x}+\frac{1}{x^{2}}\right)  \exp(-1/x),\\
G_{0}\left(  x\right)    & =\frac{\pi}{2}\exp\left(  -x/2\right)  I_{0}\left(
x/2\right)  ,\\
G_{2}\left(  x\right)    & =\frac{\pi}{4}\exp\left(  -x/2\right)  \left(
I_{0}\left(  x/2\right)  +I_{1}\left(  x/2\right)  \right)  ,\\
G_{3}\left(  x\right)    & =\frac{1}{2x}\exp\left(  -x\right)  +\frac{1}%
{2}\left(  1-\frac{1}{2x}\right)  \sqrt{\frac{\pi}{x}}\operatorname{erf}%
\left(  \sqrt{x}\right)  ,
\end{align*}
where $I_{0}\left(  x\right)$ and $I_{1}\left(  x\right)$ are
modified Bessel functions of the first kind. The approximate
solution (\ref{Jint-ap}) allows one to obtain parameter-dependent
asymptotics of (\ref{intensity}), (\ref{lifetime}). For $x\gg1$ one
gets
\begin{equation}
J\left(  x\right)  \approx\frac{\pi}{6x^{3}}\left[  \frac{\alpha^{6}}{\left(
2\beta\right)  ^{3}}+\frac{3}{4}\frac{\left(  2\beta\right)  ^{3}}{\alpha^{4}%
}-\frac{2}{3}\alpha^{3}\right]  .\label{Jint-ap1}%
\end{equation}
Since $\alpha\sim\beta^{1/2}\sim0.1$ one can put $J\left(  x\right)
\approx\left(  \pi/6\right)  \left(  \alpha^{2}/\left(  2\beta\right)
\right)  ^{3}/x^{3}$ when $x\gg1$.

It is interesting to note that, according to (\ref{intensity}) and (\ref{Jint-ap1}),
at high temperature the luminescence intensity $I\propto n/T$. The same
dependence can be obtained by assuming that the intensity of the exciton
luminescence is proportional to the number $N_{0}$ of excitons with zero
momentum \cite{Litt}, $I\propto N_{0}=\exp\left(  T_{c}/T\right)  -1\propto
n/T$ at $T\gg T_{c}=2\pi\hbar^{2}n/m$.

\section{Conclusion}We have suggested that optical
recombination of free-moving 2D excitons is generally induced by the
exciton-acoustic phonon interaction. In particular, this interaction
eliminates the uncertainty of the $Z$-component of the photon momentum.
The dependence of the exciton radiative lifetime $\tau_{R}$ on the
effective temperature $T$ of the exciton system has been derived. At
$T>T^{\ast}\approx3.6$ K the function $\tau_{R}(T)$ exhibits a
well-known linear dependence. At $T=T^{\ast}$ the lifetime
$\tau_{R}$ reaches its minimal value and at $T<T^{\ast}$ it
increases with decreasing $T$. The characteristic temperature
$T^{\ast}$ is determined only by the exciton Bohr radius and the sound
velocity in the quantum well. For the experimental verification of
the predicted non-monotonic dependence $\tau_{R}(T)$ at low
temperatures one needs nearly resonant exciton pumping and the
lattice temperature $T_{0}\ll T^{\ast}$.

The author thanks L. A. Maksimov for helpful discussions, and J. Waldie,
who has read the manuscript and made many useful remarks on its style.

\section{Appendix}According to Refs.\cite{Andr, Litt} the total optical
recombination rate $\Gamma(p)$ due to the 2D exciton - 3D
photon interaction reads (as before, we omit $\hbar$ in momentum
$\hbar k$, restoring the dimensionality only in the final results)
$\Gamma(p)=\Gamma_{1}(p)+\Gamma_{2}(p)$,
\begin{equation}
\Gamma_{1}(p)=\frac{k}{\tau_{0}\sqrt{k^{2}-p^{2}}},\text{ \ }\Gamma
_{2}(p)=\frac{\sqrt{k^{2}-p^{2}}}{\tau_{0}k},\label{Gamma-p}%
\end{equation}
where $\mathbf{p}$ is in-plane (X-Y) momentum of a 2D exciton,
$\mathbf{k}$ is the 3D photon wave vector ($k=\omega_{0}/c$), and
$\tau_{0}$ is now the radiative lifetime of an exciton with $p=0$
(note that $\tau_{0}$ in the main text has a different meaning).
$\Gamma_{1}$ and $\Gamma_{2}$ correspond to different polarizations
of the emitted photon.

Let us first reproduce the result of Ref.\cite{Andr} by finding the
average (thermal) radiative lifetime $\tau$ of an exciton if the
exciton system has a Maxwell distribution function over momenta,
$f_{p}=\exp\left(  -(E_{p} -\mu)/T\right)  $, $E_{p}=p^{2}/2m$, with
$\left\vert \mu\right\vert \gg T$.
Then the exciton density is $n=\frac{1}{S}%
{\displaystyle\sum\limits_{\mathbf{p}}}
f_{p}=mT\exp\left(  \mu/T\right)  /\left(  2\pi\hbar^{2}\right)$,
where $S$ is the X-Y plane area. In the latter formula we omit the spin
degeneracy factor (=4) of the excitons. Defining the average
intensity $I$ of the exciton luminescence as
\begin{equation}
I=n/\tau=\frac{1}{S}%
{\displaystyle\sum\limits_{\left\vert \mathbf{p}\right\vert <k}}
\Gamma(p)f_{p}\label{Int1}%
\end{equation}
and substituting $f_{p}=\exp\left(  -p^{2}/\left(  2mT\right)
\right) \left(  \frac{2\pi\hbar^{2}n}{mT}\right)$ and $
{\displaystyle\sum\limits_{\left\vert \mathbf{p}\right\vert <k}}
(...)\approx\frac{S}{4\pi\hbar^{2}}{\displaystyle\int\limits_{0}^{k^{2}}}
dp^{2}(...)$, one obtains
\begin{equation}
I=\frac{na}{\tau_{0}}%
{\displaystyle\int\limits_{0}^{1}}
dt\exp\left(  -at\right)  \frac{\left(  2-t\right)  }{\sqrt{1-t}},
\end{equation}
with $a=k^{2}/\left(2mT\right)$. Evaluating the integral, we finally
arrive at
\begin{equation}
I=\frac{n}{\tau_{0}}\left[  1+\sqrt{\pi}e^{-a}\left(
a-\frac{1}{2}\right)
F_{1}(a)\right]  ,\label{Int-final}%
\end{equation}
where $F_{1}(a)\equiv\frac{2}{\sqrt{\pi}}%
{\displaystyle\int\limits_{0}^{1}}
\exp\left(  at^{2}\right)  dt$.

\begin{figure}[!t]
\includegraphics[width=8.7cm]{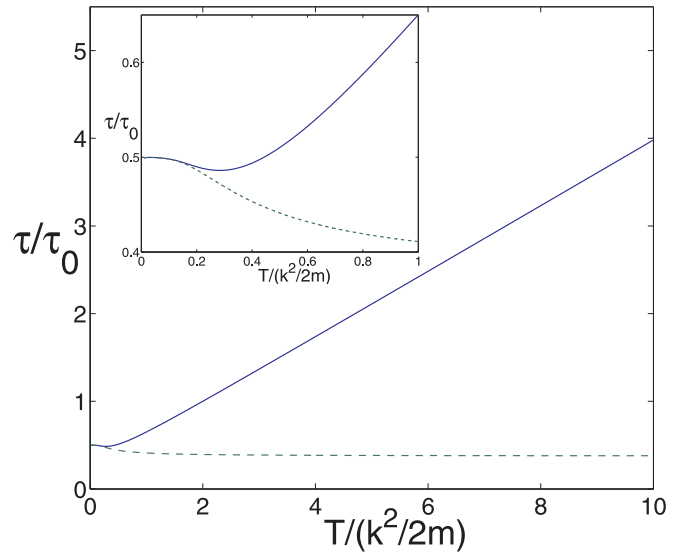}\caption{Radiative exciton lifetime $\tau$
 given by Eqs. (\ref{Int1}), (\ref{Int-final}) as a function of the effective
 exciton temperature $T$ (solid line). The dash line is $\tau (T)$
 after the replacement of $n$ by $\tilde{n}$ in the left-hand side of Eq. (\ref{Int1}).
 The inset is enlarged initial region of the main plot.}
\label{Fig3}%
\end{figure}

The dependence $\tau (T)$ given by Eqs. (\ref{Int1}) and
(\ref{Int-final}) is shown in Fig. \ref{Fig3}. At $a\ll1$, i.e., at
$T\gg k^{2}/\left( 2m\right)$ one has
$F_{1}(a)\approx2/\sqrt{\pi}+2a/\left( 3\sqrt{\pi}\right)$ and
$I\approx\frac{8}{3}\frac{n}{\tau_{0}}a=\frac{4}{3}\frac{n}{\tau_{0}}%
\frac{k^{2}}{mT}$. Substituting this result in Eq.(\ref{Int1}), we find%
\begin{equation}
\tau\approx\frac{3}{4}\tau_{0}\frac{mT}{\hbar^{2}k^{2}}\propto T.\label{Andr}%
\end{equation}
Formula (\ref{Andr}) reproduces the well-known result of Ref.
\cite{Andr}. Note that the original result is four times larger
due to the spin degeneracy factor.

At low temperature $T<\hbar^{2}k^{2}/\left( 2m\right) \sim1$ K the
dependence $\tau (T)$ exhibits a small finite upturn (inset in Fig.
\ref{Fig3}). However, the upturn might be an artefact since, on the
one hand, the temperature at the minimum is below $T_{c}$ starting
with relatively low exciton densities $n\gtrsim 10^{8}$ cm$^{-2}$
and, on the other hand, at very low density of excitons their
thermalization is hindered. In Ref.\cite{Litt}, where the authors
considered the Bose-Einstein statistics of the excitons, it was shown
that the low-temperature minimum in $\tau (T)$ is absent and the
curve monotonically comes to $0.5\tau_{0}$ (in Ref.\cite{Litt} it is
four times larger due to the spin degeneracy factor) at $T=0$.

One should note that the definition of the average radiative
lifetime of an exciton through Eq. (\ref{Int1}) used so far may
require some revision. In fact, as it is supposed, only optically active excitons
(i.e., those with momenta $\left\vert \mathbf{p}\right\vert <k$)
can contribute to the luminescence intensity. However, this has been taken
into account only in the right-hand side of Eq. (\ref{Int1}) (cp.
with Eq. (\ref{Int})). For consistency, one may replace
the total density $n$ of excitons by the density of optically active
excitons $\tilde{n}=\frac{1}{S}{\displaystyle\sum\limits_{\left\vert
\mathbf{p}\right\vert <k}} f_{p}=n\left(  1-\exp(-a)\right)$ in the
left-hand side of Eq. (\ref{Int1}).

This replacement seems well-grounded when the depletion of the
excitons with $\left\vert \mathbf{p}\right\vert <k$ due to their
optical recombination is compensated not by the relaxation of
excitons with large momenta ($\left\vert \mathbf{p}\right\vert >k$)
but by the arrival of already thermalized excitons with $\left\vert
\mathbf{p}\right\vert <k$ due to the exciton pumping. Such a situation
is possible with stationary resonant pumping which is balanced by the
recombination.

At $a\ll1$ one has $\tilde{n}\approx na$. Then equating the left and
the right sides in Eq. (\ref{Int1}) one gets
$\tilde{n}/\tau\approx\left(  8/3\right) \left( na/\tau_{0}\right)$
and $\tau\approx\left(  3/8\right)  \tau_{0}$. The next term over
$a$ brings the temperature dependence (dash line in Fig.
\ref{Fig3}). Finally,%
\begin{equation}
\tau\approx\frac{3}{8}\tau_{0}\left(
1+\frac{\hbar^{2}k^{2}}{20mT}\right)
,\label{Andr-corr}%
\end{equation}
where $T\gg\hbar^{2}k^{2}/\left( 2m\right)$. It can be seen that in the
previous result (\ref{Andr}) the large factor
$mT/\left(\hbar^{2}k^{2}\right) \gg1$ appears because of the
one-sided inclusion of all excitons as optically active ones, whereas
at high temperatures the fraction of optically active excitons is very small ($\tilde{n}/n\sim
a\ll1$).

The dependence $\tau (T)$ given by Eq. (\ref{Andr-corr}) differs
qualitatively from the experimental one \cite{Feld, Pastor}, i.e.,
theoretical $\tau$ decreases with $T$ while in the experiments $\tau
\propto T$. The discrepancy possibly comes from the pulsed pumping
used in the experiments \cite{Feld, Pastor}, where the thermal
relaxation timescale was much smaller than the pulse repetition
period. Then the relaxation can cause a rapid re-distribution of the
exciton momenta after each act of recombination and the replacement
of $n$ by $\tilde{n}$ in Eq. (\ref{Int1}) is not valid.


\begin{thebibliography}{11}
\bibitem {add}At fixed $\mathbf{k}$ the value of $q_{z}$ is also fixed by (\ref{condop}). So
at large $p\sim\sqrt{2mT}$ one has large $q_{\parallel}$ and the
main contribution is from deformation potential ($\left\vert
M_{q}\right\vert ^{2}\propto q$) whereas at small $T$ the values of
$q_{\parallel}$ are also small so the main contribution can be from
piezoelectric coupling (then $\left\vert M_{q}\right\vert
^{2}\propto q^{-\gamma}$, $\gamma\geqslant1$).
\end{thebibliography}
\end{document}